\documentclass[aps,prb,twocolumn,showpacs,citeautoscript,reprint,longbibliography]{revtex4-1}
\usepackage{graphicx}
\usepackage{bm,amsmath,amssymb,wasysym,amsfonts,dcolumn}
\usepackage[colorlinks=true, urlcolor=blue, linkcolor=blue, citecolor=blue, pdfborder={0 0 0}]{hyperref}
\usepackage[usenames,dvipsnames]{xcolor}
\usepackage{mathrsfs}

\newcommand{\be}{\begin{equation}}
\newcommand{\ee}{\end{equation}}
\newcommand{\bea}{\begin{eqnarray}}
\newcommand{\eea}{\end{eqnarray}}

\begin{document}

\title{Spin accumulation and dissipation excited by an ultrafast laser pulse}
\author{Wen-Tian Lu}
\author{Zhe Yuan}
\email[Corresponding author: ]{zyuan@bnu.edu.cn}
\affiliation{Center for Advanced Quantum Studies and Department of Physics, Beijing Normal University, Beijing 100875, China}

\date{\today}

\begin{abstract}
An ultrafast spin current can be induced by femtosecond laser excitation in a ferromagnetic (FM) thin film in contact with a nonmagnetic (NM) metal. The propagation of an ultrafast spin current into NM metal has recently been found in experiments to generate transient spin accumulation. Unlike spin accumulation in equilibrium NM metals that occurs due to spin transport at the Fermi energy, transient spin accumulation involves highly nonequilibrium hot electrons well above the Fermi level. To date, the diffusion and dissipation of this transient spin accumulation has not been well studied. Using the superdiffusive spin transport model, we demonstrate how spin accumulation is generated in NM metals after laser excitation in an FM$|$NM bilayer. The spin accumulation shows an exponential decay from the FM$|$NM interface, with the decay length increasing to the maximum value and then decreasing until saturation. By analyzing the ultrafast dynamics of laser-excited hot electrons, the ``effective mean free path'', which can be characterized by the averaged product of the group velocity and lifetime of hot electrons, is found to play a key role. The interface reflectivity has little influence on the spin accumulation in NM metals. Our calculated results are in qualitative agreement with recent experiments.
\end{abstract}

\maketitle

\section{Introduction}\label{chap1}

Since the discovery of femtosecond laser-induced ultrafast demagnetization in nickel observed by the magneto-optical Kerr experiment (MOKE) \cite{beaurepaire1996ultrafast}, various related physical phenomena have attracted much attention and have been extensively studied. The physical mechanism of ultrafast demagnetization is under debate for two possibilities: the local dissipation of angular momentum \cite{krauss2009ultrafast,carpene2008dynamics,tveten2015electron,koopmans2005unifying,koopmans2010explaining,qaiumzadeh2013manipulation,freimuth2021laser,zhang2000laser,bigot2009coherent} and nonlocal spin transport \cite{battiato2010superdiffusive,battiato2012theory,nenno2016boltzmann,nenno2018particle}. Beaurepaire {\it et al.} \cite{beaurepaire1996ultrafast} explained ultrafast demagnetization by considering the energy coupling between electrons, spins, and phonon baths via the so-called three-temperature model. Later, several mechanisms based on electron-electron \cite{krauss2009ultrafast}, electron-magnon \cite{carpene2008dynamics,tveten2015electron}, electron-phonon \cite{koopmans2005unifying,koopmans2010explaining,qaiumzadeh2013manipulation}, and electron-photon \cite{zhang2000laser,bigot2009coherent,freimuth2021laser} interactions were proposed. Unlike the local spin-flip picture, Battiato {\it et al.} \cite{battiato2010superdiffusive,battiato2012theory} proposed that the superdiffusive spin transport arising from the spin-dependent hot electrons excited by a laser pulse played a major role, which was subsequently found to have an application for broad-band terahertz emission \cite{seifert2016efficient, zhang2017bursts}.

The study of hot electron transport can be traced back to 1987, when Brorson {\it et al.} \cite{brorson1987Femtosecond} observed ultrafast electron transport after femtosecond laser excitation of a gold thin film. In particular, the velocity of the heat transport was found to be the same order of magnitude as the Fermi velocity of Au, i.e., $\sim10^6$~m/s, indicating that the heat transport was not due to electron-phonon relaxation but occurred due to the transport of nonequilibrium hot electrons well above the Fermi energy. Malinowski {\it et al.} \cite{malinowski2008control} first demonstrated that the transport of spin-polarized hot electrons was associated with ultrafast demagnetization in a magnetic multilayer excited by a laser pulse. This pioneering work stimulated many related studies in spintronics involving ultrafast hot electron transport in the following decade. An FM$|$NM bilayer was found to serve as a terahertz emitter \cite{kampfrath2013terahertz,feng2021spintronic}, in which a picosecond pulse of the spin-polarized current due to the excited hot electrons from the FM metal was injected into the NM metal. This picosecond time scale was intrinsically determined because of the difference in the spin-dependent velocity of the hot electrons. Owing to the large spin-orbit coupling of the NM metal, the spin-polarized current was then converted into a transverse charge current via the inverse spin Hall effect (ISHE). Then, the picosecond pulse of the transverse charge current produced electromagnetic radiation, whose frequency was naturally located in the terahertz region. Eschenlohr {\it et al.} did not directly excite the FM metal using the femtosecond laser. Instead, the NM metal Au in contact with FM Ni was excited and ultrafast demagnetization was observed in the Ni induced by the hot electrons that were transported into the FM metal \cite{eschenlohr2013ultrafast}. This experiment demonstrated that the direct absorption of a femtosecond laser was not the necessary condition for ultrafast demagnetization. Rudolf {\it et al.} found in a Ni$|$Ru$|$Fe trilayer with parallel magnetization for the two FM metals that laser-induced demagnetization in Ni transiently enhanced the magnetization of the Fe layer, indicating spin transfer carried by hot electrons \cite{rudolf2012ultrafast}. Later, a spin transfer arising from femtosecond laser-induced spin-polarized hot electrons was found to exert a torque on an FM metal, driving its magnetization dynamics within a subpicosecond timescale \cite{schellekens2014ultrafast}. Such ultrafast magnetization dynamics can even be realized without using an FM metal as the spin current source. For example, in a Pt/Au/GdFeCo multilayer, Wilson {\it et al.} made laser excitation in Pt and demonstrated that the ferrimagnetic GdFeCo layer was switched by injected hot electrons \cite{wilson2012ultrafast}. These findings suggest that hot electron transport is not only important in the fundamental physics of ultrafast spin transport and dynamics but also has potential application in femtosecond spintronics devices \cite{bai2020near}.

There is not a consensus in the understanding of ultrafast hot electron transport. Melnikov {\it et al.} noted that there could be a competition of ballistic and diffusive propagation for hot electron transport, and the measured spin relaxation time of the hot electrons was approximately 1~ps \cite{melnikov2011ultrafast}. Later, they experimentally demonstrated that the ultrashort spin current pulses in Fe$|$Au$|$Fe epitaxial multilayers were related to the existence of a nonthermal spin-dependent Seebeck effect corresponding to ballistic transport of spin-polarized electrons in Au \cite{alekhin2017femtosecond}. Further evidence was provided by an extremely long decay length on the order of 100~nm in Au for nonthermalized electrons \cite{alekhin2019magneto}, which were excited optically in the adjacent Fe thin film and injected through the Fe$|$Au interface as a spin filter. Bergeard {\it et al.} reported a linear dependence of the onset of demagnetization time on the thickness of the attached Cu layer and attributed the observation to the ballistic transport of hot electrons \cite{bergeard2016hot}. On the other hand, Hofherr {\it et al.} found that the demagnetization dynamics of Ni$|$Au were mainly caused by optically generated spin currents, which exhibited a transition from ballistic to diffusive transport \cite{hofherr2017speed}. B{\"u}hlmann {\it et al.} observed that the decay time of the spin polarization injected from a laser-induced ferromagnet into a thin Au layer increased with Au thickness \cite{buhlmann2020detection}. Salvatella {\it et al.} presented that the characteristic demagnetization times as a function of the Al thickness had a discontinuity at $d_{\text{Al}}$=30 nm, indicating diffusive heat transport of hot electrons \cite{salvatella2016ultrafast}.

Transient spin accumulation was observed in Cu, Ag, Pt, and Au during the laser-induced ultrafast demagnetization of an attached FM multilayer \cite{choi2014kerr,choi2014spin,choi2018magneto}. Such transient spin accumulation is significantly different from the steady-state spin accumulation in equilibrium NM metals \cite{kimura2005estimation} that is usually produced by spin pumping \cite{tserkovnyak2002,saitoh2006conversion}, the spin Hall effect \cite{sinova2015spin} and the spin Seebeck effect \cite{uchida2010spin}. Transient spin accumulation involves highly nonequilibrium hot electrons and is expected to have a very short decay time or length. However, the dynamics of ultrafast spin accumulation and spin relaxation remain unclear \cite{malinowski2018hot}. In addition, it is unknown to what extent the well-developed spin diffusion theory responsible for the steady-state spin current is still applicable for the ultrafast, photoexcited spin current \cite{ko2020optical}.

\begin{figure}[t]
  \centering
  \includegraphics[width=\columnwidth]{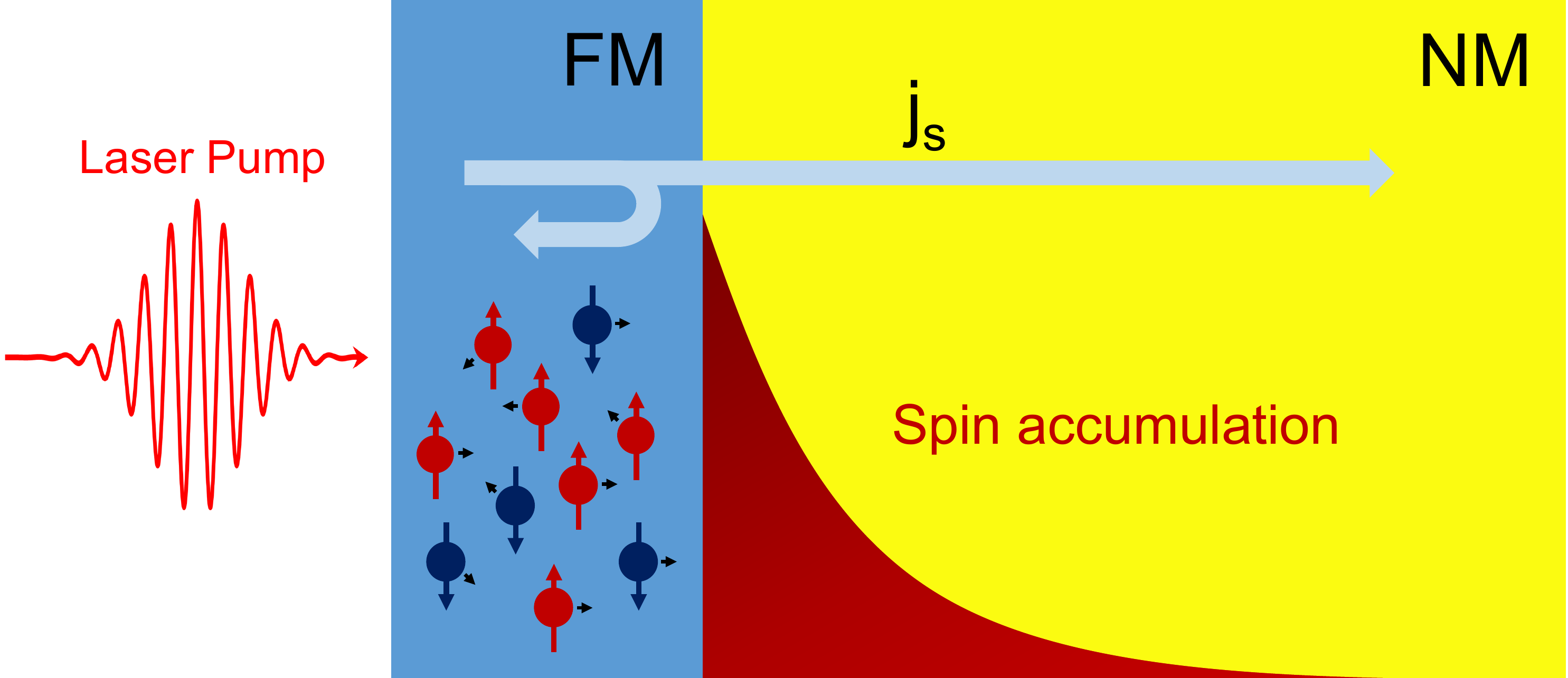}
  \caption{Schematic diagram of ultrafast spin transport and accumulation in an FM$|$NM bilayer. The hot electrons in the FM metal are excited by a femtosecond laser pulse. Owing to the spin-dependent mobility and interface reflection, an ultrafast spin current is injected into the NM layer. The elastic and inelastic scattering of the ultrafast spin current results in spin accumulation in the NM metal, which is illustrated by the red shadow.}
  \label{f1}
\end{figure}
In this article, we apply the superdiffusive spin transport theory and systematically study spin accumulation in an Fe$|$NM bilayer excited by a femtosecond laser pulse; see Fig.~\ref{f1}. Au, Al and Pt are selected as typical examples of NM metals. The spin accumulation is found to exponentially decay from the Fe$|$NM interface, and the extracted spin diffusion length $l_{s}$ exhibits a nonmonotonic dependence on time: increasing to the maximum value and then decreasing slightly until saturation. We find that this behavior is mainly caused by the movement of the center of gravity of the excited hot electrons inside the NM metal. By combining the numerical calculation and analytical analysis, we demonstrate that both the saturated value of $l_s$ and the time to reach its maximum are proportional to the ``effective mean free path'' of the hot electrons in the NM metal. The influence of interface reflection on spin accumulation is also discussed. The rest of this paper is organized as follows. The theoretical methods and computational details are briefly introduced in Sec.~\ref{chap2}. The calculated results for spin accumulation in Fe$|$NM bilayers are presented in Sec.~\ref{chap3}. This is followed by discussions about $l_s$ for its nonmonotonic variation in time (Sec.~\ref{chap3.1}), the NM metal dependence (Sec.~\ref{chap3.2}) and the effect of interface reflection (Sec.~\ref{chap3.3}). A comparison with recent experiments is provided in Sec.~\ref{chap3.4}. Conclusions are given in Sec.~\ref{chap4}.

\section{Theoretical methods and computational details}\label{chap2}

The superdiffusive spin transport model has been successfully applied to describe the experimental observation of ultrafast demagnetization resulting from excitation by a femtosecond laser pulse. In this theory, a laser pulse creates many nonequilibrium hot electrons, which are excited from the occupied states below the Fermi energy $E_F$ to the energy bands above $E_F$. The nonequilibrium electron characteristics, e.g., group velocities and lifetimes, depend on their spins in magnetic metals. For example, a photon with an energy of 1.5~eV can excite the majority-spin 3$d$ electrons in Fe to the unoccupied 4$s$ band, while the minority-spin 3$d$ electrons are excited to other unoccupied 3$d$ bands. Therefore, the excited nonequilibrium electrons in Fe have spin-dependent group velocities leading to a spin-polarized current density. Meanwhile, there are holes in the 3$d$ bands below $E_F$ that are neglected in the superdiffusive spin transport model because of their relatively low mobility \cite{ma2019plasmon}.

Nonequilibrium hot electrons above $E_F$ propagate with their own group velocity and recombine with the holes below $E_F$ during propagation due to a finite lifetime. Moreover, hot electrons are scattered by phonons, impurities, and/or other electrons. Here, we simultaneously consider the elastic scattering that conserves the scattered electron energy while changing their momentum and the inelastic scattering that allows the scattered electrons to lose energy. The latter may transfer the released energy via electron-hole recombination and excite another electron to a higher energy, which produces a cascade of electrons that contribute to transport.

The laser-excited hot electrons can transport across an FM$|$NM interface and enter the attached NM metal. Then, the spin angular momentum is transferred from the FM material to enhance the ultrafast demagnetization. Inside the NM metal, the injected nonequilibrium spin-polarized hot electrons are scattered, which leads to spin accumulation. In this work, we explicitly calculate the time- and spatial-dependent spin accumulation in NM metals after laser excitation in FM metal, while numerical techniques can be found in the previously published literature~\cite{battiato2012theory,battiato2014treating,lu2020interface}.

Since the laser spot is much larger than the mean free path of the excited hot electrons, transport can be reduced to one dimension along the interface normal of the metallic multilayers ($z$-axis). The key equation in the superdiffusive spin transport model reads \cite{battiato2012theory}
\be
\frac{\partial n_{\sigma}(E,z,t)}{\partial t} + \frac{n_{\sigma}(E,z,t)}{\tau_{\sigma}(E,z)} = \left( -\frac{\partial}{\partial z} \hat{\phi}+\hat{I} \right) S^{\rm eff}_{\sigma}(E,z,t),\label{eq:key}
\ee
where $n_{\sigma}(E,z,t)$ is the density of nonequilibrium hot electrons with spin $\sigma$ at energy $E$, position $z$ and time $t$. $\tau_{\sigma}(E,z)$ is the lifetime of hot electrons with spin $\sigma$ at energy $E$, and depends on the material via its position $z$. Thus, the second term on the left-hand side represents the dissipation of the nonequilibrium hot electrons. $S_{\sigma}^{\text{eff}}(E,z,t)$ in Eq.~\eqref{eq:key} is the effective source term for hot electrons, including scattered and newly excited electrons. $\hat{\phi}$ and $\hat{I}$ are the electron flux and identity operators, respectively. The electron flux operator $\hat{\phi}$ acting on the source term $S(z, t)$ can be explicitly expressed as
\be
\hat{\phi} S(z,t) = \int^{+\infty}_{-\infty} dz_{0} \int^{t}_{-\infty} dt_{0}\ S(z_0,t_0)\phi(z,t|z_0,t_0),
\ee
where the spin and energy indices are omitted for simplicity. The flux kernel $\phi(z, t|z_0, t_0)$ represents the electron density flux at a given position $z$ and time $t$ resulting from an electron that is excited at $z_0$ and $t_0$. The superdiffusive spin transport equation~\eqref{eq:key} is nonlocal in space and time and is solved iteratively \cite{battiato2014treating}.

Spin accumulation is defined by the difference in the majority- and minority-spin electron density, which includes the hot electrons above the Fermi energy and the occupied electrons below $E_F$, i.e.,
\begin{eqnarray}
M(z, t) &=& \int dE \left[n_{\uparrow}(E, z, t) - n_{\downarrow}(E, z, t)\right] \nonumber\\
&&+n^{\rm occ}_{\uparrow}(z, t)-n^{\rm occ}_{\downarrow}(z, t). \label{eq:m}
\end{eqnarray}
Here, $n^{\rm occ}_{\sigma}(z, t)$ is the electron density below the Fermi level with spin $\sigma$, which is the summation of the electrons that are not excited and the electrons that drop to a lower energy than $E_F$ due to inelastic scattering.

In this work, we consider three types of FM$|$NM bilayers, namely, Fe$|$Au, Fe$|$Al and Fe$|$Pt, where the thickness of Fe is always fixed at 10~nm. The thickness of the NM metals is set as 1500~nm, 600~nm, and 200~nm for Au, Al and Pt, respectively. This is to maintain a NM metal thickness that is much larger than the length scale of the spin accumulation to avoid the finite thickness effect. The spin-dependent lifetimes and velocities of excited electrons are determined using first-principles many-body calculations~\cite{zhukov2005gw+, zhukov2006lifetimes}. The excitation laser pulse is modeled using a Gaussian function with a wavelength of 780~nm (1.5~eV) and a full width at half maximum (FWHM) of 50~fs. An equal number of spin-up and spin-down electrons in Fe are excited by the laser pulse, as in the previous calculation \cite{battiato2012theory}. The nonequilibrium hot electrons between $E_F$ and $E_F+1.5$~eV are discretized in energy with an interval of 0.125~eV. The transition probability of electron scattering is set to be the same as that in the literature \cite{battiato2012theory}. Numerically, a spatial grid of 1~nm and a time step of 1~fs are employed.

It is important to note that we do not explicitly include phonons in the calculation, but phonon scattering is implicitly imposed in the elastic scattering rate to vary the momenta of the nonequilibrium hot electrons \cite{battiato2012theory}. In addition, the spin-phonon relaxation time and electron-phonon relaxation time are on the order of several picoseconds \cite{koopmans2010explaining, bigot2000electron}, which is longer than or comparable to the total time of our calculation. Therefore, electron thermalization due to electron-phonon and spin-phonon interactions is neglected. This approximation is valid unless the magnetization dynamics on a larger timescale need to be studied.

\section{Results and discussions}\label{chap3}

\begin{figure}[t]
  \centering
  \includegraphics[width=\columnwidth]{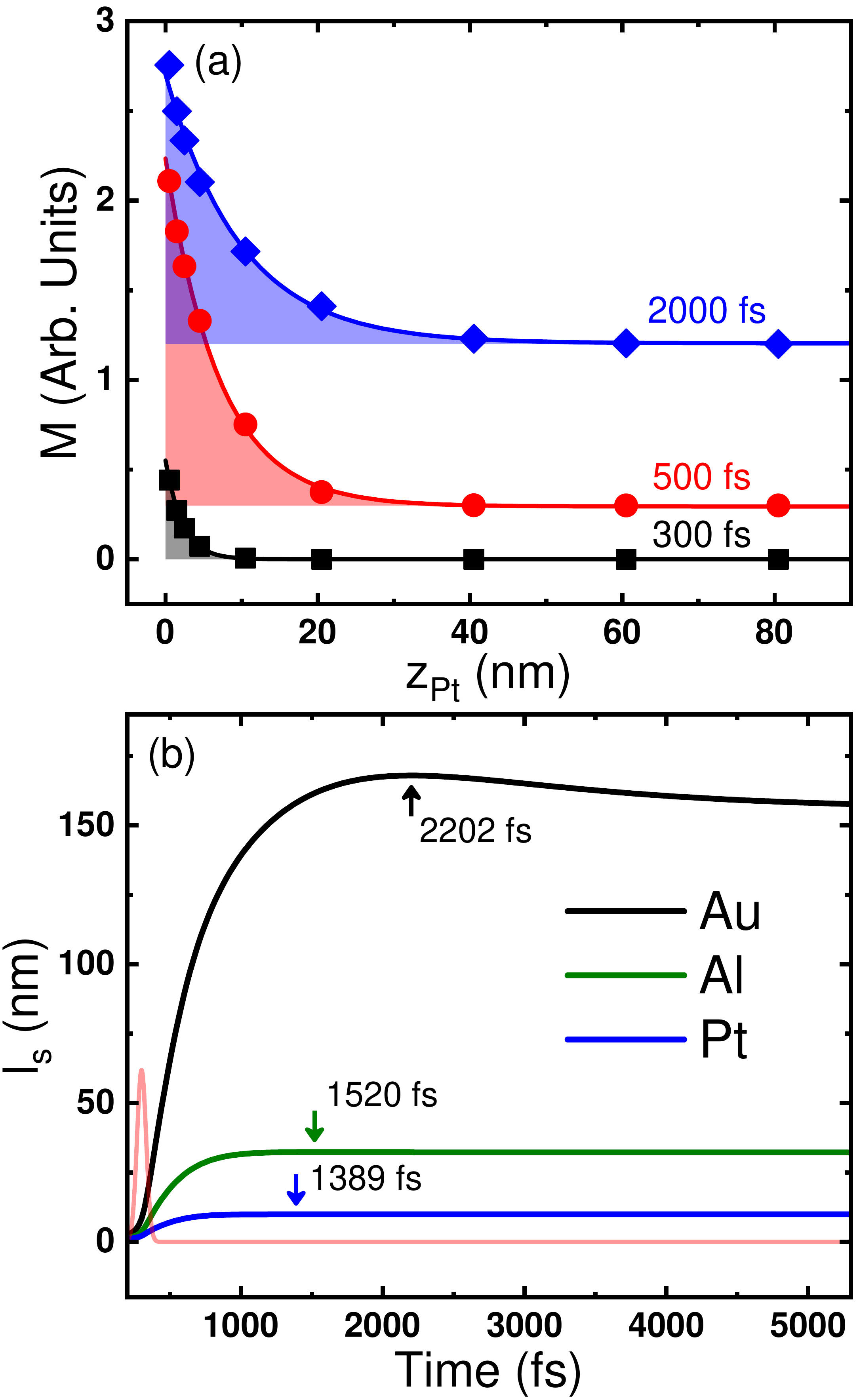}
  \caption{(a) Calculated spin accumulation in Pt at several times after excitation by a laser pulse in Fe. $z=0$ is defined at the Fe$|$Au interface. The solid lines are the exponential fits to the calculated spin accumulation to extract the effective spin diffusion length $l_s$. The data are offset for clarity. (b) The extracted $l_{s}$ for the three NM elements (NM$=$Au, Al, and Pt) in the Fe$|$NM bilayer as a function of time. The arrows denote the time at which $l_s$ reaches its maximum value. The light red line illustrates the profile of the laser pulse.}
  \label{f2}
\end{figure}
Using the superdiffusive spin transport model, the magnetization dynamics in the Fe$|$NM bilayer are calculated immediately after excitation by a femtosecond laser pulse, which has the maximum amplitude at $t=300$~fs. As an example, the spin accumulation that is defined by Eq.~(\ref{eq:m}) in Pt is plotted at several times, $t=300$, $500$, and $2000$~fs in Fig.~\ref{f2}(a). At every $t$, the calculated spin accumulation decreases from the Fe$|$Pt interface, which can be reproduced by an exponential decay, as shown by the solid lines. Analogous to the semiclassical diffusion theory of a spin current \cite{valet1993theory}, we can define the decay length as the effective ``spin diffusion length'' $l_s$.

At $t=300$~fs, the laser pulse just reaches the maximum value, and only a small part of the spin angular momentum carried by the excited hot electrons has transferred into Pt. The spin accumulation in Pt is relatively weak, and the fitted $l_s$ is short. At $t=500$~fs, hot electrons are sufficiently excited and transported across the Fe$|$Pt interface, resulting in a significant increase in spin accumulation in Pt; see the red symbols in Fig.~\ref{f2}(a). At $t=2000$~fs, the spin accumulation near the Fe$|$Pt interface decreases, while increasing in the interior Pt at $z_{\rm Pt}=10\sim50$~nm. This is because an increasing number of hot electrons move into Pt from the excitation side and are transported to a deeper position.

To better understand the behavior of spin accumulation in the NM layer, we plot the extracted spin diffusion length $l_{s}$ as a function of time in Fig.~\ref{f2}(b). The profile of the Gaussian laser pulse with an FWHM of 50~fs is also plotted as a light red line for reference. For all three NM metals, $l_{s}$ increases rapidly immediately after the excitation of the laser pulse. Then, $l_s$ reaches the maximum value at a certain time $t_{\rm max}$, as denoted by the arrows at 2202~fs (Au), 1520~fs (Al), and 1389~fs (Pt). Next, $l_{s}$ slightly decreases and saturates to a constant value in the calculated range of time. This nonmonotonic behavior is universal for the three NM metals, and among them, Au is found to have the largest saturation value of $l_s$ and the latest $t_{\rm max}$. The maximum value and saturated value of $l_s$ in Al are both less than the corresponding values in Au. This is different from the conventional spin-flip diffusion length for steady-state spin transport near the Fermi level, where due to the large spin-orbit interaction, Au would have a shorter spin-flip diffusion length than Al \cite{bass2007spin}. On the other hand, the saturated values of $l_s$ are still much shorter than the conventional spin diffusion length because the calculated time is only several picoseconds. This time scale is much less than the time needed to bring the excited system back to equilibrium, which is usually hundreds of picoseconds or nanoseconds \cite{malinowski2018hot}.

\subsection{Nonmonotonic behavior of $l_s$}\label{chap3.1}

The unexpected nonmonotonic dependence of $l_{s}$ on time is analyzed by separating the contributions from the hot electrons above $E_F$ and the occupied electronic states below the Fermi energy, i.e., the first and the second lines in Eq.~(\ref{eq:m}). Since the laser does not excite the electrons in the NM metal, the occupied electrons have equal numbers of majority and minority spins at a small $t$. After the excited hot electrons from Fe move into the NM metal, the occupied electrons are excited by the energy released from the inelastic scattering of hot electrons leaving holes below $E_F$ in the NM metal. These holes can then be recombined with hot electrons, and spin accumulation will occur because there are more hot electrons with majority spin than minority-spin electrons.

\begin{figure}[t]
  \centering
  \includegraphics[width=\columnwidth]{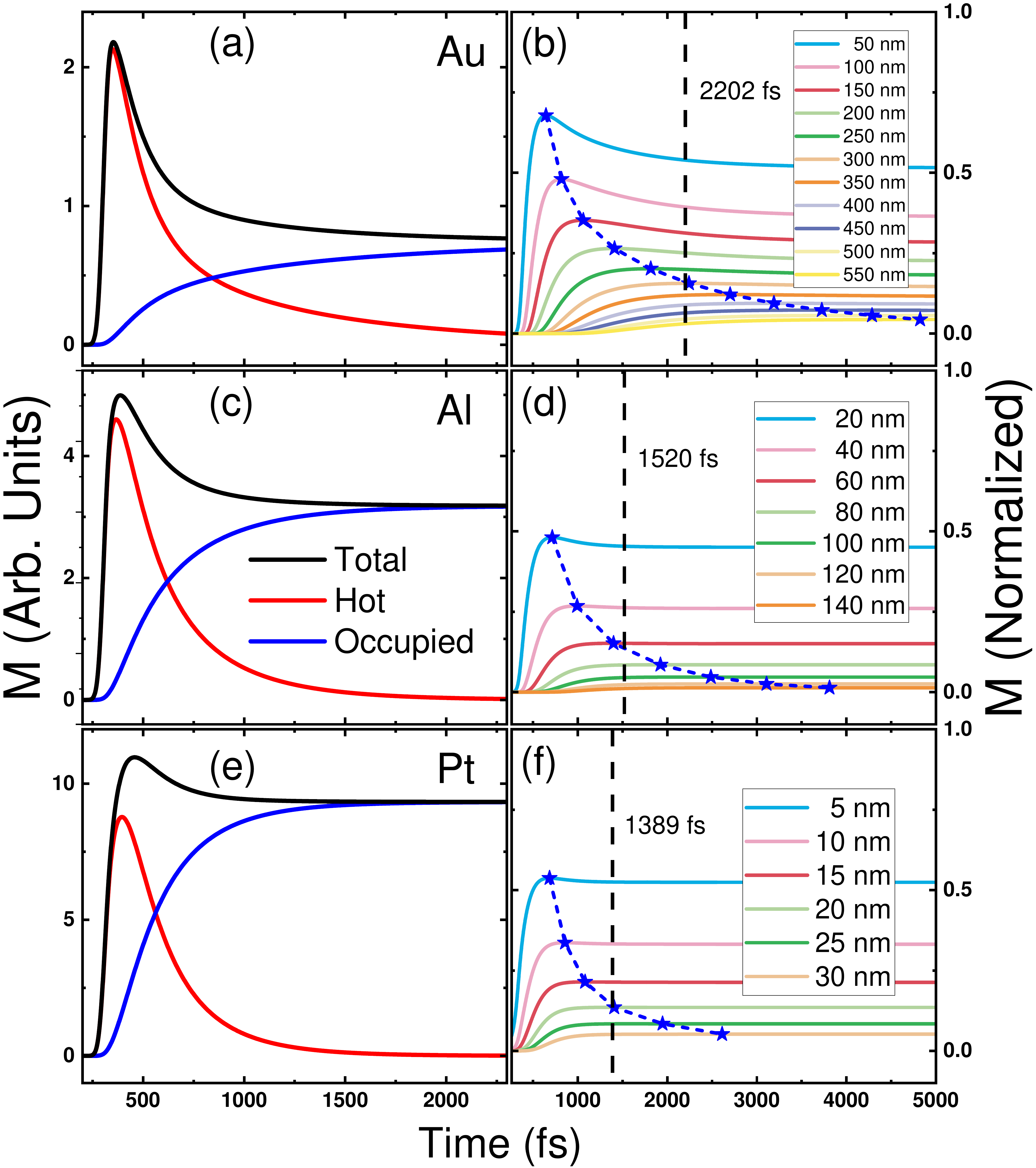}
  \caption{Calculated spin accumulation as a function of time for NM=Au [(a) and (b)], Al [(c) and (d)], and Pt [(e) and (f)]. (a), (c) and (e) show the calculated data at the Fe$|$NM interface, i.e., $z=0$. The black, red and blue lines represent the total spin accumulation, the contribution from hot electrons above $E_F$ and from the occupied state below $E_F$, respectively. (b), (d) and (f) show the normalized total spin accumulation at different positions inside the NM metal. The stars illustrate the maximum value in every curve. The vertical dashed line represents the time $t_{\rm max}$ when the maximum $l_s$ occurs, as marked by the arrows in Fig.~\ref{f2}(b).}
  \label{f3}
\end{figure}
The calculated time-dependent spin accumulation at the Fe$|$NM interface is plotted in Fig.~\ref{f3} (a), (c), and (e) for NM=Au, Al and Pt, respectively. The black, red and blue lines represent the total spin accumulation, the contribution from hot electrons above $E_F$, and the contribution from the occupied states below $E_F$, respectively. After laser excitation, the spin accumulation increases rapidly due to the excited majority-spin hot electrons that are quickly injected into the NM metal. Spin accumulation soon reaches the maximum value and then decreases slowly. The hot electrons have a relatively short lifetime, so the red lines decay within 1~ps. Meanwhile, hot electrons lose their energy to other electrons via inelastic scattering and drop below $E_F$. Then, the spin density that occurs as a result of the occupied states increases gradually and tends to saturate; see the blue line. The decay time is determined by the lifetime of the hot electrons. Au has the slowest decay time due to the longest lifetime of hot electrons in Au, while the decay time is the fastest in Pt due to its shortest lifetime. The amplitude of the spin accumulation is larger in Pt than in Au or Al because the group velocity of hot electrons in Pt is small and the spin density cannot move quickly either into the interior Pt or back to Fe. In contrast, the hot electrons in Au have a large velocity, and the spin accumulation at the Fe$|$Au interface has a small amplitude.

After understanding the time-dependent spin accumulation at the Fe$|$NM interface, we plot the normalized spin accumulation at different positions in the NM metal in Fig.~\ref{f3}(b), (d) and (f). At every time, the value at $z=0$ is normalized to be unity. Then, the overall feature at every position $z$ is very similar to the profile at $z=0$, while the maximum value is reached at a later time than at the Fe$|$NM interface, as indicated by the stars. With increasing $z$, the maximum value of the spin accumulation decreases, and the time to reach the maximum value is also delayed. This is consistent with the physical picture that spin-polarized hot electrons are injected into the NM metal and their center of gravity moves towards the interior of the NM metal with a finite rate of dissipation.

Based on the calculated curves in Fig.~\ref{f3}(b), (d) and (f), we can understand the nonmonotonic $l_s$ shown in Fig.~\ref{f2}(b) as follows. At small $t$, the spin accumulation near the Fe$|$NM interface rapidly increases to the maximum value, while at a large $z$ only increases slightly. The large difference in spin accumulation near and far from the interface results in a relatively small $l_s$. At an intermediate time, the spin accumulation near the interface decreases and tends to saturate, whereas that far from the interface approaches the maximum. The relatively small difference leads to a larger $l_s$. Long after laser excitation, the spin accumulation near the interface has already saturated, and $M$ away from the interface also decreases and tends to saturate. Then, the position-dependent difference in $M$ increases again, resulting in a reduced $l_s$. The time for the maximum $l_s$ to occur is denoted by the vertical dashed lines in Fig.~\ref{f3}(b), (d) and (f). All the vertical lines are located at the time with $M$ near the interface already tending to saturate, while $M$ in the interior NM metal has not yet reached its maximum. The location confirms the above explanation.

\subsection{Material dependence}\label{chap3.2}
The nonmonotonic time dependence of $l_{s}$ is universal for all three NM metals, but both the saturated value of $l_s$ in the long time limit and the time $t_{\rm max}$ when the maximum of $l_s$ occurs depend on the specific material. To understand the difference in the three representative NM metals, we first analytically derive the dynamics of the first generation hot electrons, i.e., the electrons excited directly by the laser pulse. In this approximation, we assume that the electrons fall below the Fermi energy as soon as they are scattered for the first time.

Because the velocities and lifetimes for both spin channels are the same in NM metals, we only consider a single spin case without loss of generality. The first-generation electron density flux $\phi_\sigma(z,t)$ as a result of electrons that are excited at position $z_0$ and time $t_0$ (or, for an NM metal, injected into the NM metal at $z=0$ and at time $t_0$) can be generally expressed as \cite{battiato2012theory}
\begin{widetext}
\begin{equation}
\phi_\sigma(z,t)=\int_{0}^{\frac{\pi}{2}}\frac{\sin\theta}{2}\exp\left[-\frac{1}{\text{cos}\theta}\int_{z_{0}}^{z}\frac{dz'}{\tau(z') v(z')}\right]\Theta\left[t-t_0-\frac{1}{\cos\theta}\int_{z_{0}}^{z}\frac{dz'}{v(z')}\right]\Theta(t-t_0)d\theta,
\end{equation}
where $\theta$ is defined as the polar angle which considers the $x$ and $y$ components of the velocity and $\Theta(t)$ is a step function with a value of 1 for $t>0$ and a value of 0 for $t\leq0$. The total electron density that has not passed through position $z$ can be obtained accordingly,
\begin{equation}
\tilde n_\sigma(z,t)=\int_{0}^{\frac{\pi}{2}}\frac{\sin\theta}{2}\left\{1-\exp\left[-\frac{1}{\text{cos}\theta}\int_{z_{0}}^{z}\frac{dz'}{\tau(z') v(z')}\right]\right\}\Theta\left[t-t_0-\frac{1}{\cos\theta}\int_{z_{0}}^{z}\frac{dz'}{v(z')}\right]\Theta(t-t_0)d\theta.
\end{equation}
Inside a homogeneous NM metal, where $v$ and $\tau$ are both constant, the above equation can be simplified as
\begin{equation}
\tilde n_\sigma(z,t)=\int_{0}^{\frac{\pi}{2}}\frac{\sin\theta}{2}\left[1-\exp\left(-\frac{z-z_0}{\tau v\cos\theta}\right)\right]\Theta\left(t-t_0-\frac{z-z_0}{v\cos\theta}\right)\Theta(t-t_0)d\theta. \label{eq4}
\end{equation}
To obtain the local electron density at position $z$, we only need to differentiate Eq.~(\ref{eq4}) with respect to position as
\begin{equation}
n_\sigma(z,t)=\frac{\partial\tilde n_\sigma(z,t)}{\partial z}=\int_{0}^{\frac{\pi}{2}}\frac{\sin\theta}{2\tau v\cos\theta}\exp\left(-\frac{z - z_0}{\tau v\cos\theta}\right)\Theta\left(t-t_0-\frac{z - z_0}{v\cos\theta}\right)\Theta(t-t_0)d\theta.
\end{equation}
\end{widetext}
The above integral can be analytically carried out in the limit of $t\rightarrow \infty$,
\begin{equation}
n_\sigma(z,t\rightarrow \infty)=\frac{1}{2 v \tau} \Gamma_0\left(\frac{z-z_0}{v \tau}\right),\label{eq6}
\end{equation}
where the upper incomplete $\Gamma$-function is defined by
\begin{equation}
\Gamma_0(x)=\int_x^\infty\frac{\exp(-t)}{t}dt.
\end{equation}

\begin{figure}[b]
  \centering
  \includegraphics[width=\columnwidth]{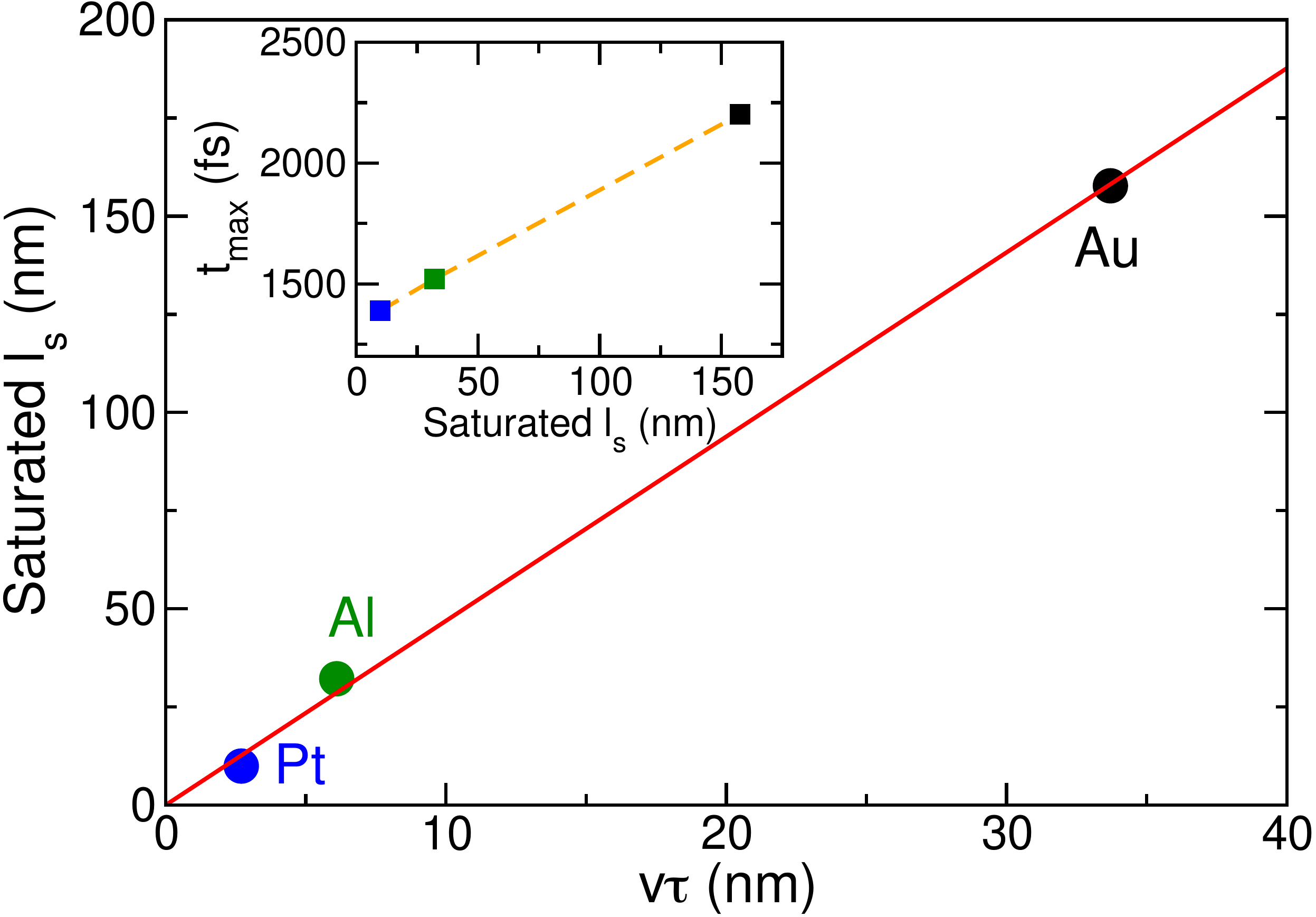}
  \caption{Saturated $l_s$ in the calculation for the three NM metals as a function of the product $v\tau$, which is averaged in the energy range from $E_F$ to 1.5~eV above $E_F$. The red solid line illustrates the proportionality. Inset: the time $t_{\rm max}$ when the maximum $l_s$ occurs as a function of the saturated $l_s$. The orange dashed line shows the linear relationship.}
  \label{f4}
\end{figure}
Note that Eq.~(\ref{eq6}) is the local distribution of the first generation electron density in the long time limit, which decays exponentially with the length scale $v\tau$. Similarly, the other spin also exhibits exponential decay with the same decay length. Therefore, the difference in the two spins, i.e., the spin accumulation $M$ at a time long after the excitation, has the same decay length $v\tau$, which determines the saturated $l_s$. To verify this analysis, we plot the saturated $l_s$ for Au, Al and Pt in Fig.~\ref{f4} as a function of the product $v\tau$ that is averaged for the hot electrons in the energy range from $E_F$ to 1.5~eV above $E_F$. $l_s$ is found to be larger than the averaged product $v\tau$, but a perfect proportionality is illustrated by the red solid line. In reality, multiple generations of excited hot electrons extend the propagation length, but scaling with the material parameter $v\tau$ still holds.

The change in spin angular momentum exhibits Elliott-Yafet-like behavior because it is only caused by the decay and reexcitation of hot electrons. The averaged product $v\tau$ can be regarded as the ``effective mean free path'' of the hot electrons. Then, the injected spin can propagate to a deeper position in an NM metal with a larger $v\tau$, as found in Fig.~\ref{f2}(b). For the time $t_{\rm max}$ when the maximum $l_s$ appears, a similar explanation is found. Near the Fe$|$NM interface, the time to reach the maximum spin accumulation is relatively fast within 500~fs despite the slight difference among the three NM metals. On the other hand, with a large ``effective mean free path'', the spin accumulation away from the interface keeps increasing and therefore postpones the time $t_{\rm max}$. Both the saturated values of $l_s$ and $t_{\rm max}$ are quantities used to characterize the propagation capability of the superdiffusive spin current, and are expected to have a positive correlation. As plotted in the inset of Fig.~\ref{f4}, the linear dependence of $t_{\rm max}$ on the saturated $l_s$ for the three NM metals is shown.

\subsection{Effect of interface reflectivity}\label{chap3.3}

At an actual Fe$|$NM interface, the potentials seen by the laser-excited hot electrons are significantly different on each side, resulting in a finite transmission probability of the incoming electrons. In particular, the transmission or reflection probability depends on the energy and spin of the hot electrons \cite{lu2020interface}. In the above calculations, we neglect the interfacial reflectivity and assume a transparent interface for all electrons. In the following, the Fe$|$Pt system is used as an example to examine the influence of the interface reflectivity on the spin accumulation and dissipation in NM metals.

\begin{figure}[t]
  \centering
  \includegraphics[width=\columnwidth]{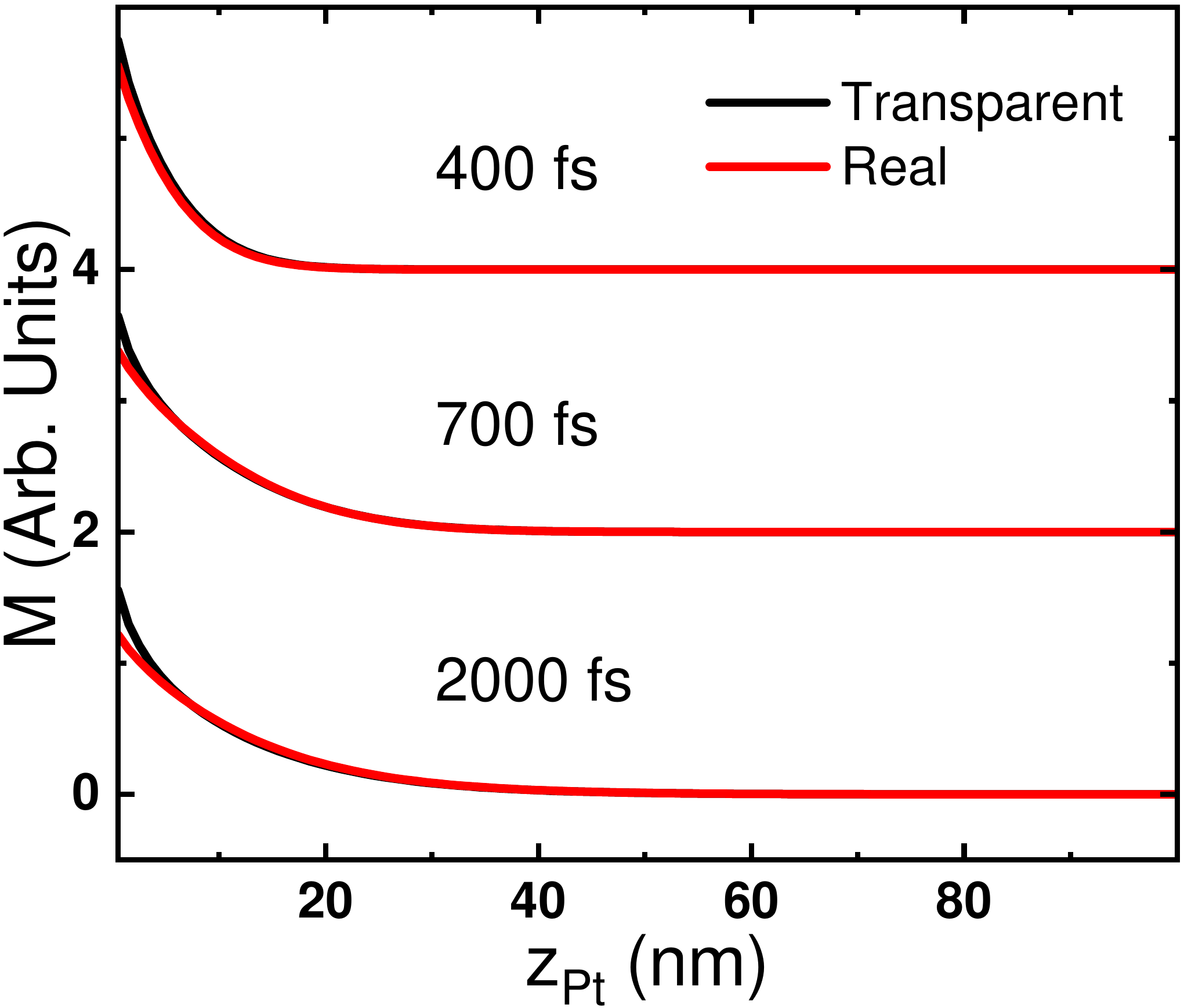}
  \caption{Calculated spin accumulation in Pt for transparent (black line) and real (red line) interfaces at $t=400$~fs, 700~fs, 2000~fs as a function of the distance from the Fe$|$Pt interface. For the real interface, the spin- and energy-dependent reflectivity obtained from first-principles transport calculations has been incorporated into the superdiffusive spin transport model \cite{lu2020interface}. The curves are offset for clarity.}
  \label{f5}
\end{figure}
Here, we follow our previous work~\cite{lu2020interface}, where the FM$|$NM interface resistance was calculated in the energy range of 0 $\sim$ 1.5 eV above the Fermi level. The spin-down electrons were found to have a higher reflectivity than the spin-up electrons, leading to the spin filtering effect in both directions. The calculated interfacial reflectivity has been incorporated into the superdiffusive spin transport model to calculate the laser-induced magnetization dynamics and terahertz emission of the Fe$|$NM bilayer \cite{lu2020interface}.

Using this scheme, we calculate the spin accumulation in Pt after laser excitation in the attached Fe layer. In Fig.~\ref{f5}, we plot the calculated spin accumulation with the real interface reflectivity (red lines) and with the transparent interface (black lines) at three times for a quantitative comparison. At $t=400$~fs, i.e., 100~fs after the maximum intensity of the laser pulse, the spin accumulations of the real and transparent interfaces are nearly identical. $M$ only has a slightly lower amplitude at the real interface. This is because of the spin filtering effect of the Fe$|$Pt interface: the spin-up electrons have a higher probability of passing through the interface than the spin-down electrons. Although the spin filtering effect increases the nonequilibrium majority spins injected into Pt, the accumulated spin at the interface also has a higher probability of flowing back to Fe. Eventually, the calculated $M$ is slightly lower only at the Fe$|$Pt interface after consideration of the real interface reflectivity.

At a later time, $t=700$~fs and $t=2000$~fs, the difference between the real and transparent interfaces becomes slightly larger but only within 5~nm from the interface. In a more interior region of Pt, the interface reflectivity does not change the numerical results of the spin accumulation. Therefore, our calculations and discussions in the previous sections are not affected by the spin filtering effect of the interface.

\subsection{Comparison with experiments}\label{chap3.4}
Our calculated results are in qualitative agreement with some recent experimental observations. By excluding optical interference, Melnikov {\it et al.} probed the spin polarization at the Au surface in the Fe$|$Au multilayer excited by a laser pulse \cite{alekhin2017femtosecond}. Since the majority-spin electrons in Fe have a higher velocity, they were injected into Au first and then minority-spin electrons were injected. Therefore, the injected spin current exhibits a positive spin polarization in the first half and a negative polarization later. However, the measured spin polarization was always positive, although it increased first and then decreased, forming a nonmonotonic behavior. The experimental phenomena are in good agreement with the calculated positive spin accumulation in the NM metal shown in Fig.~\ref{f2} and Fig.~\ref{f3}, which is mainly contributed to by the majority-spin hot electrons because the mobility of the minority-spin carriers is much weaker and usually only has a minor effect on spin accumulation. The estimated decay length of hot electrons in Au is approximately 100~nm \cite{alekhin2019magneto}, comparable with the saturated value of $l_s=157.7$~nm in our calculation. In addition, the measured decay time of the spin polarization in the Au layer injected from the attached Fe in the experiment increases with the thickness of the Au layer, suggesting that hot electron transport is not purely ballistic in Au \cite{buhlmann2020detection}. In our calculation, we could not reproduce the quantitative $l_s$ using the first-generation hot electrons due to the significant contribution of the cascade of electrons, indicating the superdiffusive nature of the ultrafast spin current. The multiple generations of hot electrons essentially exhibit a transition from ballistic to diffusive transport \cite{battiato2012theory}. Note that we only consider the electronic and magnetization dynamics in a few picoseconds, which is much smaller than the time needed for the system to return to equilibrium. To study the process for $l_s$ recovering the spin-flip diffusion length in equilibrium, a simulation must be performed with several orders of magnitude longer time and include additional relaxation mechanisms \cite{hofherr2017speed}, such as the electron-phonon and spin-phonon interaction.

\section{Conclusions}\label{chap4}

We systematically studied the spin accumulation and dissipation in NM metals with NM=Au, Al and Pt using the superdiffusive spin transport model for Fe$|$NM bilayers. The spin accumulation shows an exponential decay with the distance from the Fe$|$NM interface, which can be characterized by an ``effective spin diffusion length'' $l_s$. Unlike the constant spin-flip diffusion length in equilibrium, the calculated $l_s$ first increases and then decreases with time. This nonmonotonic behavior can be understood by analyzing the hot electron dynamics excited by a laser pulse. Specifically, the ultrafast spin current carried by hot electrons is injected from the Fe$|$NM interface, and gradually decay while moving towards the interior region in the NM metal. The competition of spin accumulation near the interface and in the interior region leads to the time-dependent variation of $l_s$. The saturated value of $l_{s}$ in the long-time limit is proportional to the averaged product of the electron velocity $v$ and lifetime $\tau$, the ``effective mean free path'' of the hot electrons. The time $t_{\rm max}$ for the maximum $l_s$ to occur is also proportional. Using Fe$|$Pt as an example, we demonstrate that the hot electron reflection at the Fe$|$NM interface only slightly reduces the spin accumulation near the interface but has very little influence on regions far from the interface. Our calculated results are in good agreement with recent experiments.

\acknowledgements
This work was partly supported by the National Natural Science Foundation of China (Grants No. 61774018, No. 12174028 and No. 11734004), the Recruitment Program of Global Youth Experts, and the Fundamental Research Funds for the Central Universities (Grants No. 2018EYT03 and No. 2018STUD03).

\end{document}